\documentclass{article}
\usepackage{fullpage}
\usepackage[american]{babel}
\usepackage{amssymb}
\usepackage{amsmath}
\usepackage{algorithmic}
\usepackage{algorithm}
\usepackage{multirow}
\usepackage{graphicx}

\newcommand{\eqn}[1]{(\ref{#1})}
\newcommand{\beql}[1]{\begin{equation}\label{#1}}
\newcommand{\eeq}{\end{equation}}
\newtheorem{theo}{Theorem}
\newtheorem{fac}{Fact}
\newtheorem{defi}{Definiton}

\title{A Practical $O(R\log\log n+n)$ time Algorithm for Computing the Longest Common Subsequence}
\author{Daxin Zhu, Lei Wang, Yingjie Wu  and Xiaodong Wang\\
}

\begin{document}
\maketitle

\begin{abstract}
In this paper, we revisit the much studied LCS problem for two given sequences. Based on the algorithm of Iliopoulos and Rahman for solving the LCS problem, we have suggested 3 new improved algorithms. We first reformulate the problem in a very succinct form. The problem LCS is abstracted to an abstract data type $DS$ on an ordered positive integer set with a special operation $Update(S,x)$. For the two input sequences $X$ and $Y$ of equal length $n$, the first improved algorithm uses a van Emde Boas tree for $DS$ and its time and space complexities are $O(R\log\log n+n)$ and $O(R)$, where $R$ is the number of matched pairs of the two input sequences. The second algorithm uses a balanced binary search tree for $DS$ and its time and space complexities are $O(R\log L+n)$ and $O(R)$, where $L$ is the length of the longest common subsequence of $X$ and $Y$. The third algorithm uses an ordered vector for $DS$ and its time and space complexities are $O(nL)$ and $O(R)$.
\end{abstract}

\section{Introduction}

The longest common subsequence (LCS) problem is a classic problem in computer science. The problem has several applications in many apparently unrelated fields ranging from file comparison, pattern matching and computational biology\cite{2,4,6,7}.

Given two sequences $X$ and $Y$, the longest common subsequence (LCS) problem is to find a subsequence of $X$ and $Y$ whose length is the longest among all common subsequences of the two given sequences.

The classic algorithm to LCS problem is the dynamic programming solution of Wagner and Fischer\cite{18}, with $O(n^2)$ worst case running time. Masek and Paterson\cite{12} improved this algorithm by using the "Four-Russians" technique to reduce its running time to $O(n^2/\log n)$ in the worst case. Since then, there has been not much improvement on the time complexity in terms of $n$ found in the literature. However, there were several algorithms with time complexities depending on other parameters. For example, Myers in\cite{13} and Nakatsu et al. in\cite{14} presented an $O(nD)$ algorithm, where the parameter $D$ is the Levenshtein distance of the two given sequences. The number of matched pairs of the two input sequences $R$, is perhaps another interesting and more relevant parameter for LCS problem. Hunt and Szymanski\cite{7} presented an $O((R+n)\log n)$ time algorithm to solve LCS problem. Their paper also cited applications, where $R\sim n$ and thus the algorithm would run in $O(n\log n)$ time for these applications. To the authors' knowledge, the most efficient algorithm sofar for solving the LCS problem is the $O(R\log\log n+n)$ time algorithm presented by Iliopoulos and Rahman\cite{8,9,10,16}. The key point of their algorithm is to solve a restricted dynamic version of the Range Maxima Query problem by using some interesting techniques of \cite{15}, and combining them with van Emde Boas structure\cite{17}.
Readers are referred to\cite{2} for a more comprehensive comparison of the well-known algorithms for LCS problem and their behavior in various application.

In this paper, we will revisit the classic LCS problem for two given sequences and present new algorithms with some interesting new observations and some novel ideas. Our main result is an improved $O(R\log\log n+n)$ time algorithm of Iliopoulos and Rahman. Our novel algorithm has a very simple structure.
It is very easy to implement and thus very practical.

The organization of the paper is as follows.

In the following 4 sections, we describe our improved $O(R\log\log n+n)$ time algorithm of Iliopoulos and Rahman for solving LCS problem.

In Section 2, the preliminary knowledge for presenting our algorithm for LCS problem is discussed, and the $O(R\log\log n+n)$ time algorithm of Iliopoulos and Rahman is reviewed briefly.
In Section 3, we present our improvements on the algorithm of Iliopoulos and Rahman with time complexity $O(R\log\log n+n)$, where $n$ and $R$ are the lengths of the two given input strings, and the number of matched pairs of the two input sequences, respectively.
Some concluding remarks are located in Section 4.

\section{An $O(R\log\log n+n)$ time Algorithm}

In this section, we briefly review the $O(R\log\log n+n)$ time algorithm of Iliopoulos and Rahman\cite{10} for the sake of completeness.

A sequence is a string of characters over an alphabet $\sum$. A subsequence of a sequence $X$ is obtained by deleting zero or more characters from $X$ (not necessarily contiguous). A substring of a sequence $X$ is a subsequence of successive characters within $X$.

For a given sequence $X=x_1x_2\cdots x_n$ of length $n$, the $i$th character of $X$ is denoted as $x_i \in \sum$ for any $i=1,\cdots,n$. A substring of $X$ from position $i$ to $j$ can be denoted as $X[i:j]=x_ix_{i+1}\cdots x_j$. If $i\neq 1$ or $j\neq n$, then the substring $X[i:j]=x_ix_{i+1}\cdots x_j$ is called a proper substring of $X$. A substring $X[i:j]=x_ix_{i+1}\cdots x_j$ is called a prefix or a suffix of $X$ if $i=1$ or $j=n$, respectively.

\begin{defi}\label{df1}

An appearance of sequence $X=x_1x_2\cdots x_n$ in sequence $Y=y_1y_2\cdots y_n$, for any $X$ and $Y$, starting at position $j$ is a sequence of strictly increasing indexes $i_1,i_2,\cdots,i_n$ such that $i_1=j$, and $X=y_{i_1},y_{i_2},\cdots,y_{i_n}$.

A match for sequences $X$ and $Y$ is a pair $(i,j)$ such that $x_i=y_j$.
The set of all matches, $M$, is defined as follows:
$$M = \{(i,j) | x_i = y_j, 1\leq i,j \leq n\}$$

The total number of matches for $X$ and $Y$ is denoted by $R=|M|$.
\end{defi}

It is obvious that $R\leq n^2$.

\begin{defi}\label{df2}

A common subsequence of the two input sequences $X=x_1x_2\cdots x_n$ and $Y=y_1y_2\cdots y_n$, denoted $cs(X,Y)$, is a subsequence common to both $X$ and $Y$. The longest common subsequence of $X$ and $Y$, denoted $LCS(X,Y)$, is a common subsequence whose length is the longest among all common subsequences of the two given sequences. The length of $LCS(X,Y)$ is denoted as $r(X,Y)$.
\end{defi}

In this paper, the two given sequences $X$ and $Y$ are assumed to be of equal length. But all the results can be easily extended to
the case of two sequences of different length.

\begin{defi}\label{df3}
Let $T(i,j)$ denote $r(X[1:i],Y[1:j])$, when $(i,j)\in M$. $T(i,j)$ can be formulated as follows:
\beql{eq21}
T(i,j)=\left\{\begin{array}{ll}
\textrm{Undefined} & \textrm{if } (i,j)\not\in M\\
1 & \textrm{if } (i,j)\in M \textrm{and }(i',j')\not\in M, i'<i,j'<j\\
\max_{
\begin{subarray}{l}
1\leq l_i<i\\
1\leq l_j<j\\
(l_i,l_j)\in M\\
\end{subarray}
} \{T(l_i,l_j)\}
 & \textrm{Otherwise }\\
\end{array} \right.
\eeq

\end{defi}

In the algorithm of Iliopoulos and Rahman\cite{5}, a vector $H$ of length $n$ is used to denote the maximum value so far of column $l$ of $T$.
For the current value of $i\in[1..n]$, $H_i(l) = max_{1\leq k<i,(k,l)\in M}\{T(k,l)\},1\leq l\leq n$.
The footnote $i$ is used to indicate that the current row number is $i$, and the values of $H_i(l),1\leq l\leq n$ are not changed for row $i$ in the algorithm. We can omit the footnote $i$ if the current row is clear.

The most important function $RMQ_i(left,right)=max_{left\leq l\leq right}\{H_i(l)\}$ used in the algorithm is a range maxima query on $H$ for the range $[left..right], 1\leq left\leq right\leq n$.
It is clear that

\beql{eq22}
T(i,j)=1+RMQ_i(1,j-1), \textbf{ if } (i,j)\in M
\eeq

For the efficient computation of $T(i,j)$, the following facts\cite{9,15} are utilized in the algorithm of Iliopoulos and Rahman.

\begin{fac}\label{fac1}
Suppose $(i,j)\in M$. Then for all $(i',j)\in M, i'>i$, (resp.$(i,j')\in M, j'>j$), we must have $T(i',j)\geq T(i,j)$ (resp.$T(i,j')\geq T(i,j)$ ).
\end{fac}

\begin{fac}\label{fac2}
The calculation of the entry $T(i,j), (i,j)\in M, 1\leq i,j\leq n$, is independent of any $T(l,q), (l,q)\in M, l=i,1\leq j\leq n$.
\end{fac}

The algorithm is proceed in a row by row manner as follows.

\begin{algorithm}
\caption{Iliopoulos-Rahman-LCS}
\begin{algorithmic}[1]
\FOR{$i=1$ to $n$}
\STATE $H \leftarrow S$  \COMMENT{Update $H$ for the next row}\\
\FORALL{$(i,j)\in M$}
\STATE $T(i,j) \leftarrow 1+RMQ(1,j-1)$\\
\STATE $S(j) \leftarrow T(i,j)$\\
\ENDFOR
\ENDFOR
\end{algorithmic}
\end{algorithm}

In the algorithm, another vector $S$, of length $n$, is invoked as a temporary vector. After calculating $T(i,j)$, the vector $S$ is restored from $T$, $S(j)=T(i,j)$, if $(i,j)\in M$. Therefore, at the end of processing of row $i$, $S$ is actually $H_{i+1}$. The algorithm continues in this way as long as it is in the same row. As soon as it comes a new row, the vector $H$ is updated with new values from $S$. In the algorithm above, $RMQ(1,j-1)$ is actually $RMQ_i(1,j-1)$. We can omit the footnote $i$ since the processing is in the same row $i$. The correctness of the above procedure follows from Facts 1 and 2. The problem $RMQ$ can be solved in $O(n)$ preprocessing time and $O(1)$ time per query\cite{3,4}. Therefore, for the constant time range maxima query, an $O(n)$ preprocessing time has to be paid as soon as $H$ is updated. Due to Fact 2, it is sufficient to perform this preprocessing once per row. So, the computational effort added for this preprocessing is $O(n^2)$ in total.

The most complicated part of the algorithm of Iliopoulos and Rahman is in the computation of $RMQ(1,j-1)$. To eliminate the $n^2$ term from the running time of the algorithm, a van Emde Boas tree is used to store the information in $H$. This data structure can support the operations $Search$, $Insert$, $Delete$, $Min$, $Max$, $Succ$, and $Pred$ in $O(\log\log n)$ time. By using these operations the query $RMQ(1,j-1)$ can then be answered in $O(\log\log n)$ time.
The $O(n)$ preprocessing step of the algorithm can then be avoided and hence the $n^2$ term can be eliminated from the running time.
However, as a price to pay, the query time of $RMQ(1,j-1)$ increases to $O(\log\log n)$.

Finally, the time complexity is improved to $O(R\log\log n+n)$, using $O(n^2)$ space.
The vEB structure described in the algorithm of Iliopoulos and Rahman is somewhat involved. For more details of the data structure, the readers are referred to\cite{9,15}.

\section{Improvements of the algorithm}
In this section, we will present several improvements on the algorithm of Iliopoulos and Rahman.

\subsection{A first improvement}
In the algorithm of Iliopoulos and Rahman, the set $M = \{(i,j) | x_i = y_j, 1\leq i,j \leq n\}$ is constructed explicitly in the lexicographic order of the matches such that all the matches can be processed in the correct order.
To this purpose, two separate lists, $L_X$ and $L_Y$ are built in $O(n)$ time. For each symbol $c\in\sum$, $L_X(c)$ (resp. $L_Y(c)$) stores, in sorted order, the positions of $c$ in $X$ (resp. $Y$), if any.
Then, a van Emde Boas structure is used to build the set $M$ in $O(R\log\log n)$ time.
This preprocessing step can be removed completely from our improved algorithm, since we do not need to access each match $(i,j)$ directly. It will be clear in the description of the algorithm in latter section.

Secondly, to avoid overlap, the information in $H$ are stored separately in row $i$ and $i+1$. We have noticed that, the only query on $H$ used in the algorithm is of the form $RMQ(1,j-1)=\max_{1\leq l<j}\{H(l)\}$. It is obvious that if in each row $i$, we process all $j\in L_Y(x_i)$ in a decreasing order, i.e. the order of $j$ from large to small, then every overlap for current $j$ will not change the values $RMQ(1,j-1)$ for the succeed $j$. Therefore, the information in $H$ can be updated in same row immediately. In our improved algorithm, the list $L_Y$ is built in this way such that for each symbol $c\in\sum$, the positions of $c$ in $Y$ are listed from large to small.

\subsection{The key improvements}

By the definition of $H_i(l) = max_{1\leq k<i,(k,l)\in M}\{T(k,l)\},1\leq l\leq n$ for the current value of $i\in[1..n]$, it is not difficult to observe the following fact.
\begin{fac}\label{fac3}
For all $1\leq i'<i\leq n$, and $1\leq j\leq n$, $H_{i'}(j)\leq H_{i}(j)$.
\end{fac}

The information in $H$ we maintained is used for computing $RMQ(1,j-1)$ in the algorithm. If we use a vector $Q$ of length $n$ to store the values of $Q(j)=RMQ(1,j)$ for all $1\leq j\leq n$, then $RMQ(1,j-1)$ will be computed more directly by $Q(j-1)$ in $O(1)$ time. For example, if $H=(0,1,2,1,2,0,1)$, then $Q=(0,1,2,2,2,2,2)$.
An important albeit easily observable fact about $Q$ is that, its components $Q(j), 1\leq j\leq n$ are nondecreasing.
\begin{fac}\label{fac4}
Suppose $Q(j)=RMQ(1,j)=max_{1\leq l\leq j}\{H(l)\}$. Then the values of $Q(j), 1\leq j\leq n$ form a nondecreasing sequence.
This sequence is under a very special form. If $Q(j)<Q(j+1), 1\leq j<n$, then $Q(j+1)=Q(j)+1$. Therefore, the values of this sequence are taken from a consecutive integer set $\{0,1,\cdots, L\}$, where $L=r(X,Y)$ is the length of the longest common subsequence of $X$ and $Y$.
\end{fac}

\noindent{\bf Proof.}

For any $1\leq j'<j\leq n$, $Q(j')=max_{1\leq l\leq j'}\{H(l)\}$ and $Q(j)=max_{1\leq l\leq j}\{H(l)\}$.
It follows from $j'<j$ that $\{l|1\leq l\leq j'\}\subset\{l|1\leq l\leq j\}$. It follows then $Q(j')\leq Q(j)$. This proves that the values of $Q(j), 1\leq j\leq n$ form a nondecreasing sequence.

In the case of $Q(j)<Q(j+1), 1\leq j<n$, since $Q(j+1)$ and $Q(j)$ are both nonnegative integers, we have:
\beql{eq31}
Q(j+1)\geq Q(j)+1
\eeq

On the other hand, by the definition of $Q$, if current row number is $i$, we have:

\beql{eq32}
\begin{split}
Q(j+1)& =\max_{1\leq l\leq j+1}\{H_i(l)\}\\
 & =\max\{\max_{1\leq l\leq j}\{H_i(l)\},H_i(j+1)\}\\
 & =\max\{Q(j),H_i(j+1)\}\\
 & =H_i(j+1)\\
\end{split}
\eeq

It follows from the definition of $H_i(j+1)$ that

\beql{eq33}
\begin{split}
H_i(j+1)& = \max_{1\leq k<i,(k,j+1)\in M}\{T(k,j+1)\}\\
 & =T(k',j+1), 1\leq k'<i,(k',j+1)\in M\\
 & =Q_{k'}(j)+1\\
\end{split}
\eeq

It follows from Fact \ref{fac3} and $k'<i$ that $H_{k'}(j)\leq H_{i}(j)$, and thus $Q_{k'}(j)\leq Q_i(j)=Q(j)$.

It follows from \eqn{eq32} and \eqn{eq33} that

\beql{eq34}
\begin{split}
Q(j+1)& =H_i(j+1)\\
 & =Q_{k'}(j)+1\\
 & \leq Q(j)+1\\
\end{split}
\eeq

Combining \eqn{eq31} and \eqn{eq34} we then have, $Q(j+1)=Q(j)+1$.

The proof is completed. $\Box$

By the definition of $Q$ we know, if current row number is $i$, then for $1\leq j\leq n$,

\beql{eq35}
\begin{split}
Q(j)& =\max_{1\leq l\leq j}\{H_i(l)\}\\
 & =\max_{1\leq l\leq j}\max_{1\leq k<i,(k,j)\in M}\{T(k,j)\}\\
 & =\max_{\begin{subarray}{l}
 1\leq k<i\\
 1\leq l\leq j\\
 \end{subarray}
} \{T(k,l)\}\\
\end{split}
\eeq

At the end of the algorithm, all the $n$ rows are treated. At this time, we have, for $1\leq j\leq n$,

\beql{eq36}
Q(j)=\max_{\begin{subarray}{l}
 1\leq k\leq n\\
 1\leq l\leq j\\
 \end{subarray}
} \{T(k,l)\}
\eeq

In other words, $Q(j)=r(X,Y[1:j])$, $1\leq j\leq n$. Especially, $Q(n)=L=r(X,Y)$.
If we want compute $L=r(X,Y)$, but not $LCS(X,Y)$, then we do not need to store the 2 dimensional array $T$. At the end of the algorithm, $Q(n)$ will return the answer.

Because of its unusual form, $Q$ can be viewed as a piecewise linear function. As we know, it is sufficient to record the break points of such functions to calculate their values. Therefore, we can use another vector $P$ of length at most $L$ to record the break points of $Q$. Let $m=\max_{1\leq i\leq n}\{Q(i)\}$. For $1\leq j\leq n$, the value of $P(j)$ can be defined as follows:

\beql{eq37}
P(j)=\left\{\begin{array}{ll}
\min_{1\leq i\leq n}\{i|Q(i)=j\} & \textrm{if } j\leq m\\
n+1 & \textrm{Otherwise }\\
\end{array} \right.
\eeq

Let

\beql{eq38}
S(j)=P(j), \textbf{  if  }1\leq j\leq n, 1\leq P(j)\leq n
\eeq


It is clear that $S$ forms an increasing sequence of length at most $L$.

For instance, if $Q=(0,1,2,2,2,2,2)$, then $P=(2,3,8,8,8,8,8)$, and $S=\{2,3\}$.

Let $\alpha=|S|$, then $S(\alpha)$ is the maximal element of $S$.

By the definition of $S$, if $k<\alpha$, then for any $S(k)\leq j<S(k+1)$, we have $Q(j)=k$. In the case of $k=\alpha$, for any $S(\alpha)\leq j\leq n$, we have $Q(j)=\alpha$. Therefore, $Q(j)$ can be easily computed by $S$ as follows.

\beql{eq39}
Q(j)=\left\{\begin{array}{ll}
k & \textrm{if } S(k)\leq j<S(k+1)\\
\alpha & j\geq S(\alpha)\\
\end{array} \right.
\eeq

Furthermore, we can wind up the following fact.

\begin{fac}\label{fac5}
Suppose $P(j),1\leq j\leq n$ be defined by formula \eqn{eq37}. Then $S=\{P(j)|1\leq j\leq n, 1\leq P(j)\leq n\}$ forms an increasing sequence of length at most $L$.
This sequence has a very unique property. For each $P(t)\in S$, and any $P(t)\leq j<P(t+1)$, $t=
\max_{\begin{subarray}{l}
 1\leq k<i\\
 1\leq l\leq j\\
 \end{subarray}
}
\{T(k,l)\}$, if current row number is $i$. At the end of the algorithm, for each $P(t)\in S$, $t=r(X,Y[1:j])$ for any $P(t)\leq j<P(t+1)$. Especially, the maximal element of $S$ is $P(L)$, and $L=r(X,Y)$.
\end{fac}

\noindent{\bf Proof.}

By the definition of $P$, we have, for each $P(t)\in S$, if $P(t)\leq j<P(t+1)$, then $Q(j)=t$. Therefore, it follows from formula \eqn{eq35} that if current row number is $i$, then
$t=\max_{\begin{subarray}{l}
 1\leq k<i\\
 1\leq l\leq j\\
 \end{subarray}
} \{T(k,l)\}$

At the end of the algorithm, all the $n$ rows are treated. At this time, for each $P(t)\in S$,
$t=\max_{\begin{subarray}{l}
 1\leq k\leq n\\
 1\leq l\leq j\\
 \end{subarray}
} \{T(k,l)\}=r(X,Y[1:j])$.

Especially, if $P(t')$ is the maximal element of $S$, then $P(t')\leq n<P(t'+1)=n+1$, and thus,

$t'=r(X,Y[1:n])=r(X,Y)=L$, i.e., the maximal element of $S$ is $P(L)$, and $L=r(X,Y)$.

The proof is completed. $\Box$

In the algorithm, if current match $(i,j)$ is processed, then the value of $T(i,j)$ is changed to $1+Q(j-1)$.
Let $Q(j')$ be the successor of $Q(j-1)$ in $Q$, i.e.,

\beql{eq310}
j'=\min_{j\leq l\leq n}\{l|Q(l)>Q(j-1)\}
\eeq

Then $Q[j:j'-1]$ must be changed to $1+Q(j-1)$, according to the definition of $Q$.

Similarly, Let $S(k)$ be the successor of $j-1$ in $S$, i.e.,

\beql{eq311}
k=\min_{1\leq l\leq n}\{l|S(l)>j-1\}
\eeq

Then $S(k)$ must be changed to $j$, according to the definition of $S$.

In the case of $j-1$ has no successor in $S$, i.e., $j-1\geq S(\alpha)$, then $j$ must be added into $S$.

For example, if $Q=(0,1,2,2,2,2,2)$, $S=(2,3)$, and current match $(4,6)$ is processed, then $T(4,6)$ is changed to $1+Q(j-1)=3$. In this case, $j=6$ and $j'=8$, and thus $Q[6:7]$ must be replaced by $3$. The current values of $Q$ becomes $Q=(0,1,2,2,2,3,3)$. At this point, $j-1=5>S(\alpha)=3$, and thus $j$ must be added to $S$. The current values of $S$ becomes $S=(2,3,6)$.

\subsection{The improved algorithm}

It is readily seen from the discussions above that if we can use the ordered set $S$ defined by formula \eqn{eq39} to calculate the function $RMQ(1,j-1)$, then the algorithm will be simplified substantially.
The key point is the way to maintain the ordered set $S$ efficiently.

Let $U=\{j|1\leq j\leq n\}$.
Suppose $DS$ be an abstract data type on an ordered positive integer set $S$. The abstract data type $DS$ can support the following operations on $S$:

\begin{enumerate}

\item $Size(S)$

A query on an ordered positive integer set $S$ that returns the number of integers in $S$.

\item $Succ(S,x)$

A query that, given a positive integer $x$ whose key is from $U$, returns the next larger integer in $S$, or $0$ if $x$ is the maximum integer in $S$.

\item $Pred(S,x)$

A query that, given a positive integer $x$ whose key is from $U$, returns the next smaller integer in $S$, or $0$ if $x$ is the minimum integer in $S$.

\item $Update(S,x)$

A modifying operation that, given a positive integer $x$, if $Succ(S,x-1)>0$, then replace $Succ(S,x-1)$ with the integer $x$, otherwise augments the set $S$ with a new integer $x$.

\end{enumerate}

With this abstract data type, we can maintain the ordered positive integer set $S\subseteq U$ defined by formula \eqn{eq38} in our new algorithm $LCS$ as follows.

\begin{algorithm}
\caption{$LCS$}
\begin{algorithmic}[1]
\STATE $S \leftarrow \emptyset$\\
\FOR{$i=1$ to $n$}
\FORALL{$j\in L_Y(x_i)$}
\STATE $Update(S,j)$\\
\ENDFOR
\ENDFOR
\RETURN $Size(S)$
\end{algorithmic}
\end{algorithm}

In above algorithm $LCS$, the list $L_Y(c), c\in \Sigma$ stores, in a decreasing order, the positions of $c$ in $Y$. The $|\Sigma|$
lists can be constructed in $O(n)$ time, by simply scanning $Y$ in turn. At the end of the algorithm, $L=r(X,Y)=Size(S)$, the length of $LCS(X,Y)$, is returned.

The efficiency of the new algorithm is depended heavily on the efficiency of the abstract data type $DS$, especially on the efficiency of its operation $Update(S,x)$.

We have found that the van Emde Boas tree is an elegant data structure for our purpose.
Specifically, van Emde Boas trees support each of the following dynamic set operations, $Search$, $Insert$, $Delete$, $Min$, $Max$, $Succ$, and $Pred$ in $O(\log\log n)$ time. The operation $Update(S,x)$ can be easily implemented by combining at most two successive operations $delete$ and $insert$.

If we chose van Emde Boas tree as our data structure for $S$, the new algorithm can be described as follows.

\begin{algorithm}
\caption{$vEB\_LCS$}
\begin{algorithmic}[1]
\STATE $S \leftarrow \emptyset$\\
\FOR{$i=1$ to $n$}
\FORALL{$j\in L_Y(x_i)$}
\STATE $k \leftarrow Succ(S,j-1)$\\
\IF {$k<Max(S)$}
\STATE $Delete(S,k)$\\
\ENDIF
\STATE $Insert(S,j)$\\
\ENDFOR
\ENDFOR
\RETURN $Size(S)$
\end{algorithmic}
\end{algorithm}

The structure of algorithm $vEB-LCS$ is very simple. Although the algorithm can correctly return the length of $LCS(X,Y)$, it does not directly give $LCS(X,Y)$.
If we want to compute $LCS(X,Y)$, but not just its length, we have to record more information. A commonly used method is to record the predecessor of each match $(i,j)\in M$ by a two dimensional array like $T(i,j)$. The two dimensional array returned by the algorithm allows us to quickly construct an $LCS$ of $X$ and $Y$. This method requires extra $O(n^2)$ space. For this purpose, we can design a more efficient method using only $O(R)$ space. We use two vectors $B$ and $C$, both of length $R$, to record the predecessor's match number and the matched character for each match $(i,j)\in M$, respectively. The match numbers for all matches $(i,j)\in M$ are generated one after another in the algorithm.

\begin{algorithm}
\caption{$vEB\_LCS$}
\begin{algorithmic}[1]
\STATE $m \leftarrow 0, S \leftarrow \emptyset$\\
\FOR{$i=1$ to $n$}
\FORALL{$j\in L_Y(x_i)$}
\STATE $k \leftarrow Succ(S,j-1)$\\
\IF {$k<Max(S)$}
\STATE $Delete(S,k)$\\
\ENDIF
\STATE $Insert(S,j)$\\
\STATE $p \leftarrow Pred(S,j)$\\
\STATE $m \leftarrow m+1$\\
\STATE $B(m) \leftarrow D(p), C(m) \leftarrow j, D(j)\leftarrow m$\\
\ENDFOR
\ENDFOR
\RETURN $Size(S)$
\end{algorithmic}
\end{algorithm}

In the above algorithm, the vector $D$ is a temporary vector of length at most $L$ used to store the match numbers for current set $S$.
With the two vectors $B$ and $C$ computed in the above algorithm, the following recursive algorithm prints out $LCS(X,Y)$. The initial call is $Print-LCS(Size(S))$.

\begin{algorithm}
\caption{$Print\_LCS(k)$}
\begin{algorithmic}[1]
\IF {$k\leq 0$}
\RETURN
\ELSE
\STATE $Print-LCS(B(k))$\\
\PRINT $Y(C(k))$\\
\ENDIF
\end{algorithmic}
\end{algorithm}

It is obvious that $Print-LCS(Size(S))$ takes time $O(L)$, since it prints one character of $LCS(X,Y)$ in each recursive call.
Finally, we can find that the following theorem holds.

\begin{theo}\label{th1}
The algorithm $vEB\_LCS$ correctly computes $LCS(X,Y)$ in $O(R\log\log n+n)$ time and $O(R)$ space in the worst case, where $R$ is the total number of matches for $X$ and $Y$.
\end{theo}

\noindent{\bf Proof.}
It follows from formulas \eqn{eq39} and \eqn{eq311} that the values of $RMQ(1,j-1)$ for each row $i$ can be correctly computed by the set $S$ maintained in the algorithm. Therefore, the algorithm can compute $LCS(X,Y)$ correctly as the original algorithm of Iliopoulos and Rahman.
It is obvious that the computation of list $L_Y(c), c\in \Sigma$, costs $O(n)$ time and space.
For each match $(i,j)\in M$, the algorithm executes each of the 4 operations $Succ$, $Succ$,$Succ$, and $Succ$ at most once, and each of the 4 operations costs $O(\log\log n)$. Therefore, the total time spent for these operations is $O(R\log\log n)$. The worst case time complexity of the algorithm is therefore $O(R\log\log n+n)$.

To maintain the van Emde Boas tree, $O(n)$ space is sufficient. To reconstruct the $LCS(X,Y)$, the algorithm uses two vectors $B$ and $C$, both of size $R$. The space required by the algorithm is thus $O(R)$. The worst case space complexity of the algorithm is therefore $O(R)$.

The proof is completed. $\Box$

If we chose a balanced binary search tree such as red-black tree as our data structure for the ordered positive integer set $S$, the following dynamic set operations, $Search$, $Insert$, $Delete$, $Min$, $Max$, $Succ$, and $Pred$ can be implemented in $O(\log |S|)$ time. In this case, The time complexity of our algorithm becomes $O(R\log L)$, since $|S|\leq L$.
We then can find that the following theorem holds.

\begin{theo}\label{th2}
The longest common subsequence problem can be solved in $O(R\log L+n)$ time and $O(R)$ space in the worst case, where $n,L$ and $R$ are the length of input sequences $X$ and $Y$, the length of $LCS(X,Y)$, and the total number of matches for $X$ and $Y$, respectively.
\end{theo}

The ordered positive integer set $S$ in our algorithm can also be efficiently purported by an ordered vector $s$ of length at most $L$ as follows.

\begin{algorithm}
\caption{$V\_LCS$}
\begin{algorithmic}[1]
\STATE $\alpha \leftarrow 0$, $k \leftarrow -1$\\
\FOR{$i=1$ to $n$}
\FORALL{$j\in L_Y(x_i)$}
\WHILE{$k\geq 0$ \AND $s(k)\geq j$}
\STATE $k \leftarrow k-1$\\
\ENDWHILE
\STATE $s(k+1)\leftarrow j$\\
\IF{$k=\alpha$}
\STATE $\alpha\leftarrow 1+\alpha$\\
\ENDIF
\STATE $k \leftarrow \alpha$\\
\ENDFOR
\ENDFOR
\RETURN $\alpha$
\end{algorithmic}
\end{algorithm}

In the above algorithm, for each row $i$, all columns $j\in L_Y(x_i)$ are processed in a decreasing order. The successors of all $j-1$ in $s$ are searched and updated also in a decreasing order from the largest element $s(\alpha)$. It is obvious that the time spent for each row $i$ is $O(\alpha)$. Therefore, we can conclude that the time complexity of the above algorithm is $O(nL)$, since $\alpha\leq L$.

\begin{theo}\label{th3}
The longest common subsequence problem can be solved in $O(nL)$ time and $O(R)$ space in the worst case, where $n$ and $L$ are the length of input sequences $X$ and $Y$, and the length of $LCS(X,Y)$, respectively.
\end{theo}

\end{document}